\documentclass[epj]{webofc}
\usepackage{txfonts}

\wocname{EPJ Web of Conferences}
\woctitle{CONF12}

\usepackage{bm}
\usepackage{mathtools,amssymb}
\def\be{\begin{equation}}
\def\ee{\end{equation}}
\newcommand*{\vx}{\bm{\mathrm{x}}}
\newcommand*{\vy}{\bm{\mathrm{y}}}
\newcommand*{\vz}{\bm{\mathrm{z}}}
\newcommand*{\vp}{\bm{\mathrm{p}}}

\newcommand*{\vA}{\bm{\mathrm{A}}}
\newcommand*{\e}{\mathop{}\!\mathrm{e}}
\newcommand*{\valpha}{\bm{\alpha}}

\DeclarePairedDelimiter{\ket}{\lvert}{\rangle}
\DeclarePairedDelimiter{\bra}{\langle}{\rvert}
\DeclarePairedDelimiter{\vev}{\langle}{\rangle}
\newcommand*{\Eqref}[1]{Eq.~\eqref{#1}}

\expandafter\let\expandafter\udot\csname\encodingdefault\string\d\endcsname
\DeclareRobustCommand*{\d}{\relax\ifmmode\mathop{}\!\mathrm{d}\else\expandafter\udot\fi}
\newcommand*{\dfr}[2][]{{\ifx&#1&\frac{\mathrm{d}#2}{2\pi}\else\frac{\mathrm{d}^{#1}#2}{(2\pi)^{#1}}\fi}}

\DeclareMathOperator{\Det}{Det}
\DeclareMathOperator{\Tr}{Tr}

\newcommand*{\calD}{\mathcal{D}}

\makeatletter
\def\I{\@ifstar\@@ImU\@ImU}
\newcommand*{\@ImU}{\mathrm{i}}
\newcommand*{\@@ImU}{\mathrm{i}\mkern1mu}
\makeatother

\begin{document}

\selectlanguage{english}
\title{Dyson--Schwinger Approach to Hamiltonian QCD}
\author{Davide Campagnari\inst{1}\fnsep\thanks{Presenter; \email{d.campagnari@uni-tuebingen.de}.} \and
        Hugo Reinhardt\inst{1} \and Markus Q. Huber\inst{2} \and Peter Vastag\inst{1} \and Ehsan Ebadati\inst{1}
}

\institute{%
Institut of Theoretical Physics, University of T\"ubingen, Auf der Morgenstelle 14, 72076 T\"ubingen, Germany
\and
Institute of Physics, University of Graz, NAWI Graz, Universit\"atsplatz 5, 8010 Graz, Austria
}

\abstract{%
Dyson--Schwinger equations are an established, powerful non-perturbative tool for QCD.
In the Hamiltonian formulation of a quantum field theory they can be used to perform
variational calculations with non-Gaussian wave functionals. By means of the DSEs
the various $n$-point functions, needed in expectation values of observables like the
Hamilton operator, can be thus expressed in terms of the variational kernels of our
trial ansatz. Equations of motion for these variational kernels are derived
by minimizing the energy density and solved numerically.
}

\maketitle

\section{Introduction}

Coulomb gauge Yang--Mills theory has been investigated in the continuum both in the
Hamiltonian \cite{Schutte:1985sd,Szczepaniak:2001rg,Feuchter:2004mk,Epple:2006hv}
and the Lagrangian \cite{Watson:2006yq,Watson:2007vc,Reinhardt:2008pr,Watson:2010cn} approach.
In the Hamiltonian approach, the use of variational methods with Gaussian wave functionals
\cite{Feuchter:2004mk,Epple:2006hv} has led to an equal-time gluon propagator which
behaves like the photon in the UV but is strongly suppressed in the IR, signalling confinement.
The obtained propagator also compares favourably with the available lattice data
\cite{Burgio:2008jr}; some deviations in the mid-momentum regime can be accounted for
by using non-Gaussian wave functionals \cite{Campagnari:2010wc}. Such functionals can be
handled by means of DSEs, exploiting the formal similarity
between vacuum expectation values in the Hamiltonian formalism and correlation functions
in Euclidean quantum field theory. The emerging DS-type of equations were dubbed canonical recursive
Dysin--Schwinger equations \cite{Campagnari:2010wc,Campagnari:2015zsa}. In this talk we report about the
derivation of the CRDSEs for full QCD and their application to the study of
the spontaneous breaking of chiral symmetry.

Chiral symmetry breaking in Coulomb gauge has been studied for example in
Refs.~\cite{Finger:1981gm,Adler:1984ri,Alkofer:1988tc}. While it has been shown that an
infrared enhanced potential can account for chiral symmetry breaking, the calculated
physical quantities, such as the dynamical mass and chiral condensate, turn out to be
far too small. A wave functional including the coupling of the quarks to the transverse
gluons improves the results towards the phenomenological findings \cite{Pak:2011wu}.
We apply here the techniques developed in Refs.~\cite{Campagnari:2010wc,Campagnari:2015zsa}
to the wave functional proposed in Ref.~\cite{Pak:2011wu} and extended in
Refs.~\cite{Vastag:2015qjd,Campagnari:2016wlt}.

\section{Hamiltonian Approach to Quantum Chromodynamics}

Canonical quantization of QCD relies usually on the temporal gauge $A_0^a=0$, and results
in a functional Schr\"odinger equation. Gauge invariance of the wave functional is enforced
by the Gauss law, which can be explicitly solved in Coulomb gauge, yielding the gauge-fixed
Hamiltonian
\be\label{neu2}
\begin{split}
H ={}& \frac12\int\d^3x \, \biggl( J_A^{-1} \frac{\delta}{\delta A_i^a(\vx)} J_A \frac\delta{\delta A_i^a(\vx)} + B_i^a(\vx)B_i^a(\vx) \biggr) \\
&+ \frac{g^2}{2} \int\d^3x \d^3y \, J_A^{-1} \, \rho^a(\vx) \, J_A \, F^{ab}_A(\vx,\vy) \,\rho^b(\vx) \\
&+ \int\d^3x \,  \hat\psi^{\dag}(\vx) \bigl[ -\I* \alpha_i \partial_i + \beta m - g \alpha_i t_a A^a_i(\vx) \bigr]  \hat\psi(\vx) .
\end{split}
\ee
Here, $\beta$ and $\alpha_i$ are the usual Dirac matrices; $t_a$ are the generators of
the gauge group in the fundamental representation (or, in general, in the representation
to which the fermion fields belong); $J_A=\Det(G_A^{-1})$ is the Faddeev--Popov determinant; and
\be\label{cx2}
\bigl[G_A^{ab}(\vx,\vy)\bigr]^{-1} = \bigl( - \delta^{ab} \partial^2 - g f^{acb} A_i^c(\vx) \partial_i \bigr) \delta(\vx-\vy)
\ee
is the Faddeev--Popov operator, with $g$ being the coupling constant and $f^{abc}$
being the structure constants of the $\mathfrak{su}(N_\mathrm{c})$ algebra. The second term
in \Eqref{neu2} represents the Coulomb-like interactions of the total colour charge
\[
\rho^a = \hat\psi^{\dag} t_a \hat\psi - \I* f^{abc} A_i^b \frac{\delta}{\delta A^c_i}
\]
through the Coulomb kernel
\be\label{neu3}
F_A \equiv G_A (-\nabla^2) G_A \, .
\ee

\section{Dyson--Schwinger Equations in the Hamiltonian Approach}

The derivation of the CRDSEs in pure Yang--Mills theory has been presented in
Ref.~\cite{Campagnari:2010wc}, and the generalization to full QCD has been given in
Ref.~\cite{Campagnari:2015zsa}. Here we summarize briefly the steps leading from the
choice of the wave functional to the corresponding CRDSEs.

A state $\ket\varPhi$ in the fermionic Fock space is conveniently represented by
coherent-states defined in terms of Grassmann fields
$\xi$ and $\xi^\dag$ \cite{Campagnari:2015zsa}
\be\label{dir3}
\langle\xi\vert\Phi\rangle \equiv \varPhi(\xi^\dag,\xi) .
\ee
The Dirac field operators $\hat\psi$ and $\hat\psi^\dag$ act onto this state as
\be\label{vev3}
\hat{\psi} \varPhi(\xi^\dag,\xi) = \left( \xi_- + \frac{\delta}{\delta \xi^\dag_+}\right) \varPhi(\xi^\dag,\xi) , \qquad
\hat{\psi}^\dag  \varPhi(\xi^\dag,\xi) = \left( \xi^\dag_+ + \frac{\delta}{\delta \xi_-} \right) \varPhi(\xi^\dag,\xi) ,
\ee
where $\xi_\pm = \Lambda_\pm \xi$, and $\Lambda_\pm$ are the projectors onto states of positive and negative energy
of the free Dirac operator. We have put explicitly a hat over the fermion operators $\hat\psi$, $\hat\psi^\dag$ in
\Eqref{vev3} to distinguish them from the classical Grassmann fields $\xi$, $\xi^\dag$
used in the coherent-state representation.

The expectation value in the vacuum state $\varPsi$ of an operator $K$ depending on
bosonic and fermionic fields is given by
\be\label{mad1}
\vev{K[A,\Pi,\hat\psi,\hat\psi^\dag]}
= \int \calD\xi^\dag \, \calD\xi  \,\calD A \, J_A \e^{-\xi^\dag(\Lambda_+-\Lambda_-)\xi} \, \varPsi^*[A,\xi^\dag,\xi] \,
K[A,\tfrac{\delta}{\I\delta A},\hat\psi,\hat\psi^\dag] \, \varPsi[A,\xi^\dag,\xi].
\ee
In \Eqref{mad1} the functional integration runs over transverse gauge field configurations
satisfying the Coulomb gauge condition, $\partial_i A_i^a=0$, and is restricted to the
first Gribov region;
the exponential factor occurring in the fermionic functional integration in \Eqref{mad1}
arises from the completeness relation of the coherent fermion states \cite{Berezin:1966nc,Campagnari:2015zsa}.
Once the fermion field operators and the gluon canonical momentum act onto the vacuum
wave functional, \Eqref{mad1} boils down to an expectation value of some functional of
the fields to be integrated over.

The vacuum state of QCD can be assumed to be of the form
\be\label{qcd0}
\varPsi[A,\xi,\xi^\dag] \eqqcolon \exp\left\{ -\frac12 \, S_A[A] - S_f[\xi,\xi^\dag,A] \right\} ,
\ee
where $S_A$ is a functional of the gauge field only, while $S_f$ contains both the
fermion and the gluon fields. The expectation value of a function
$f(A,\xi,\xi^\dag)$ of the fields is a path integral of the form
\[
\vev{F} = \int\calD A \, \calD\xi^\dag \, \calD\xi \, J_A \, \e^{-\xi^\dag(\Lambda_+-\Lambda_-)\xi} \e^{-S_A - S_f - S_f^*} \, f(A,\xi,\xi^\dag)
\]
which looks like a correlation function of a Euclidean field theory with ``action''
\[
S = S_A + S_f + S_f^* + \xi^\dag(\Lambda_+-\Lambda_-)\xi -\Tr\ln G_A^{-1}
\]
Correspondingly, we can derive Dyson--Schwinger equations for the gluon and quark
correlators by starting from the identity
\be\label{cx3}
0 = \int \calD A \, \calD\xi^\dag \, \calD\xi \: \frac{\delta}{\delta\phi}
\left\{ J_A \, \e^{-\xi^\dag(\Lambda_+-\Lambda_-)\xi} \:
\varPsi^*[A,\xi,\xi^\dag] \, K[A,\xi,\xi^\dag] \, \varPsi[A,\xi,\xi^\dag] \right\}
\ee
with $\phi\in\{A,\xi,\xi^\dag\}$, while ghost DSEs can be derived by starting from the
operator identity \Eqref{cx2}. (It may be useful to introduce ghost fields in the Hamiltonian
approach but this is not strictly necessary.)

The ``Dyson--Schwinger'' equations derived from \Eqref{cx3} are not equations of motion
in the usual sense, but rather relate the Green functions of the theory to the (so far
undetermined) vacuum wave functional. This is why we dubbed them canonical recursive
Dyson--Schwinger equations (CRDSEs).

\section{The Vacuum Wave Functional}

The explicit form of the vacuum wave functional is unknown. We will therefore solve the 
Schr\"odinger equation in an approximate fashion by means of the \emph{variational principle}:
we take an ansatz for the wave functional, depending on some variational kernels, we evaluate
the expectation value of the Hamilton operator \Eqref{neu2}, thereby using the CRDSEs,
and minimize the resulting vacuum energy density with respect to the variational kernels.

To proceed further, we need an explicit ansatz for the vacuum wave functional \Eqref{qcd0}.
For the Yang--Mills part we choose an ``action'' $S_A$ involving up to quartic terms in
the gluon field
\be\label{cx4}
S_A = \omega(1,2) \, A(1) \, A(2) + \frac{1}{3!} \, \gamma_3(1,2,3) \,  A(1) \, A(2) \, A(3)
+ \frac{1}{4!} \, \gamma_4(1,2,3,4) \,  A(1) \, A(2) \, A(3) \, A(4) ,
\ee
where $\omega$, $\gamma_3$, and $\gamma_4$ are variational kernels.
We are using here a compact notation where a repeated numerical label denotes integration
over the spatial coordinates as well as summation over discrete indices (colour, Lorentz\ldots).
For the quark wave functional we choose the ansatz \cite{Pak:2011wu,Vastag:2015qjd,Campagnari:2016wlt}
\[
S_f = \xi_+^\dag(1) \, K_A(1,2) \, \xi_-(2),
\qquad \xi_{\pm}=\Lambda_{\pm}\xi.
\]
The kernel $K_A$ contains both a purely fermionic part and the coupling of the quarks
to the transverse gluons
\be\label{cx6}
K_A^{mn}(\vx,\vy) = \delta^{mn} \beta \, s(\vx,\vy) + g \, t_a^{mn}
\int \d^3z \, \bigl[ v(\vx,\vy;\vz) \, \alpha_i + w(\vx,\vy;\vz) \, \beta \, \alpha_i \bigr] A^a_i(\vz) .
\ee
Here, the functions $s$, $v$ and $w$ are the variational kernels.
Note that the quark-gluon coupling in \Eqref{cx6} contains besides the leading-order
term $\sim\alpha_i$ known from perturbation theory \cite{Campagnari:2014hda} also a second
Dirac structure $\sim\beta\alpha_i$, which turns out to be of fully non-perturbative nature.

\begin{figure}[t]
\parbox{.45\linewidth}{\includegraphics[width=\linewidth]{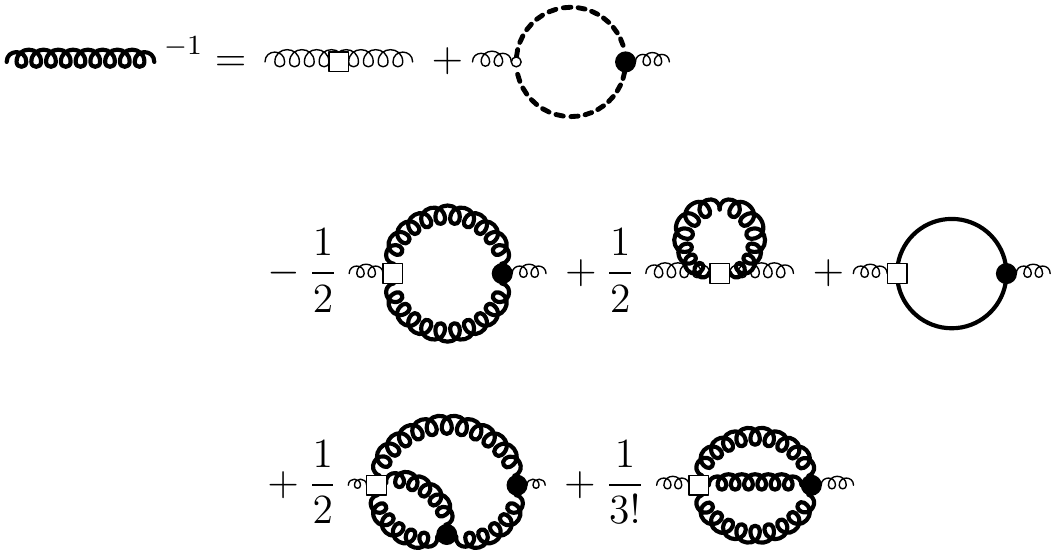}}%
\hfill
\parbox{.45\linewidth}{\centering%
\includegraphics[width=\linewidth]{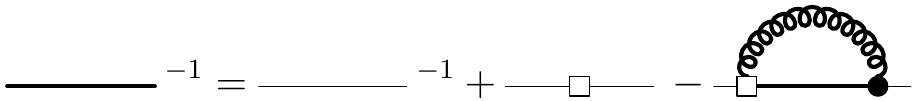}\par\addvspace{6ex}\includegraphics[width=.8\linewidth]{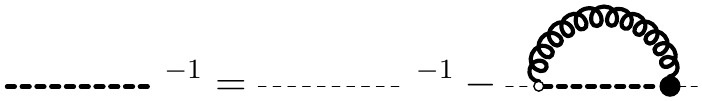}}
\caption{\label{fig-props}CRDSEs for the gluon propagator (left),
quark propagator (top right), and ghost propagator (bottom right). Empty boxes stand for
the variational kernels, fat dots for the full (one-particle irreducible) vertices, and
fat lines for fully dressed propagators.}
\end{figure}

With the ansatz given by Eqs.~\eqref{cx4}--\eqref{cx6} we find e.g.~the propagator equations
represented diagrammatically in Fig.~\ref{fig-props}.

\section{Yang--Mills Correlation Functions}

\begin{figure}[t]
\centering
\parbox{.4\linewidth}{\centering\includegraphics[width=\linewidth]{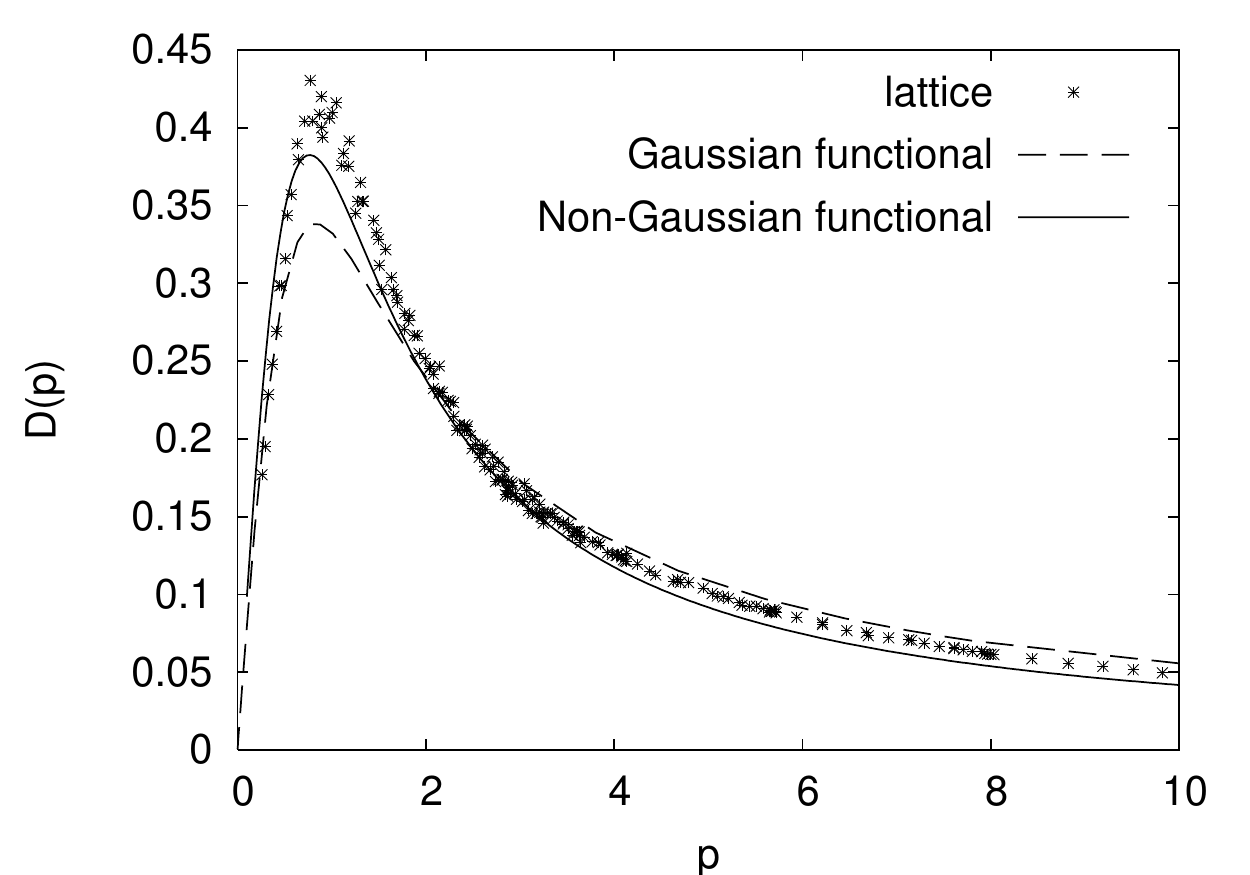}\\(a)}%
\qquad
\parbox{.4\linewidth}{\centering\includegraphics[width=\linewidth]{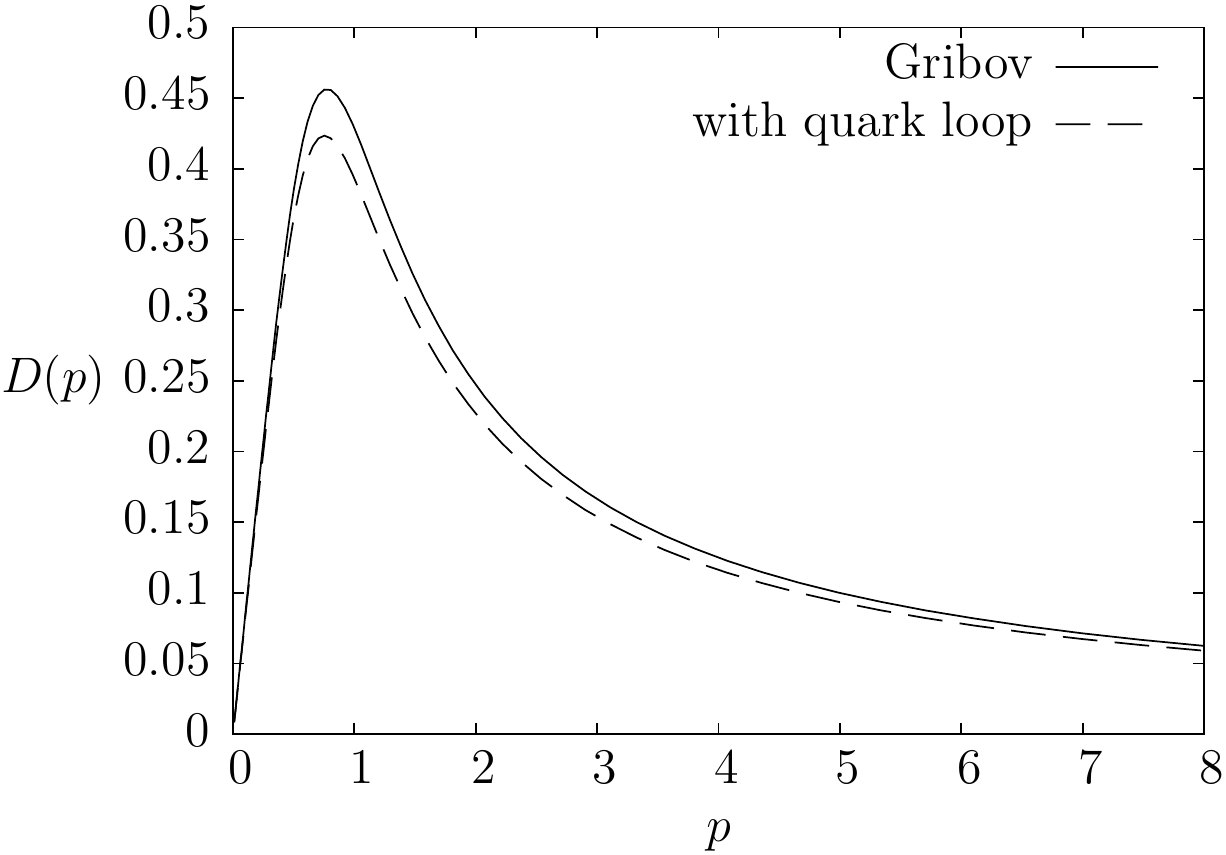}\\(b)}
\caption{\label{fig-glp}(a) Quenched gluon propagator with Gaussian and non-Gaussian wave functional.
(b) Gluon propagator with one dynamical quark family.}
\end{figure}

The first investigations in the Hamiltonian approach \cite{Feuchter:2004mk,Epple:2006hv,Epple:2007ut}
used a Gaussian-type wave functional,
which allows to evaluate the expectation value of the Hamilton operator in terms of
products of propagators by means of Wick's theorem. While the obtained equal-time gluon
propagator agrees with lattice simulations in the IR and in the UV, there is some
mismatch in the mid-momentum regime. Figure~\ref{fig-glp}a shows the results for the
gluon propagator with a non-Gaussian functional \cite{Campagnari:2010wc}, showing that
the presence of a three-gluon coupling (and hence of a gluon loop in the CRDSE) improves
the agreement with the lattice data. Figure~\ref{fig-glp}b illustrates the effect of the
quark loop onto the gluon propagator.

\begin{figure}[ht]
\centering
\parbox{.4\linewidth}{\centering\includegraphics[width=\linewidth]{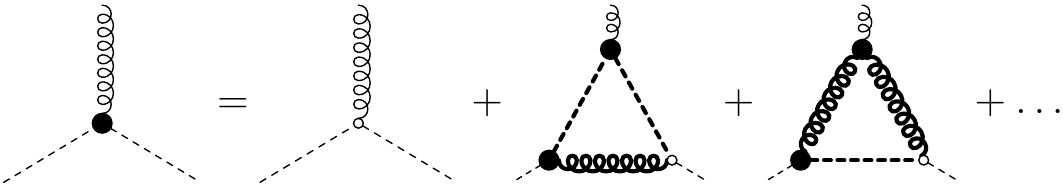}\\[1ex](a)\\[3ex]\includegraphics[width=.9\linewidth]{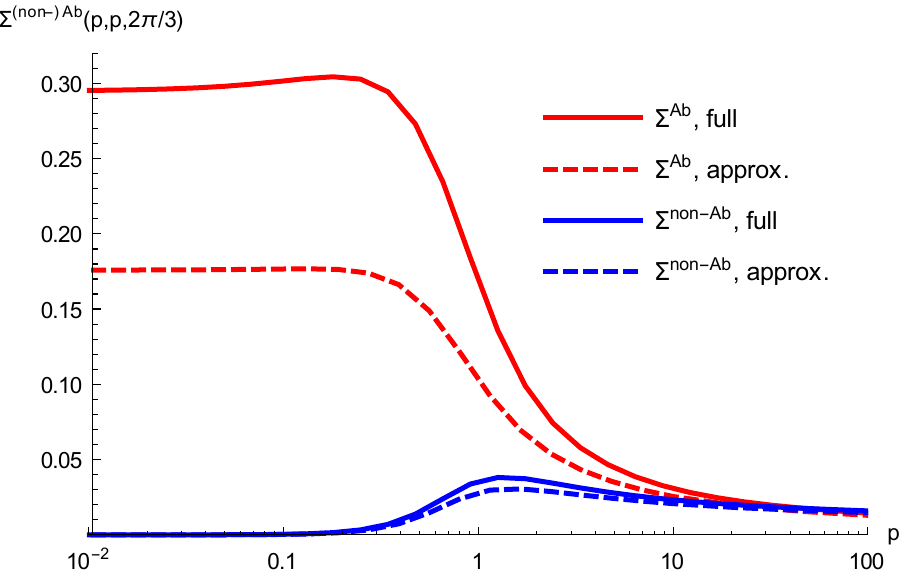}\\(c)}%
\qquad\qquad
\parbox{.4\linewidth}{\centering\includegraphics[width=\linewidth]{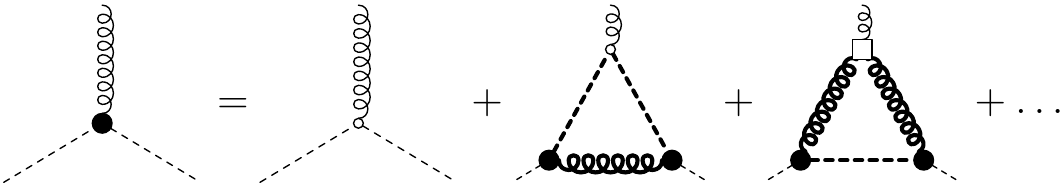}\\[1ex](b)\\[3ex]\includegraphics[width=.9\linewidth]{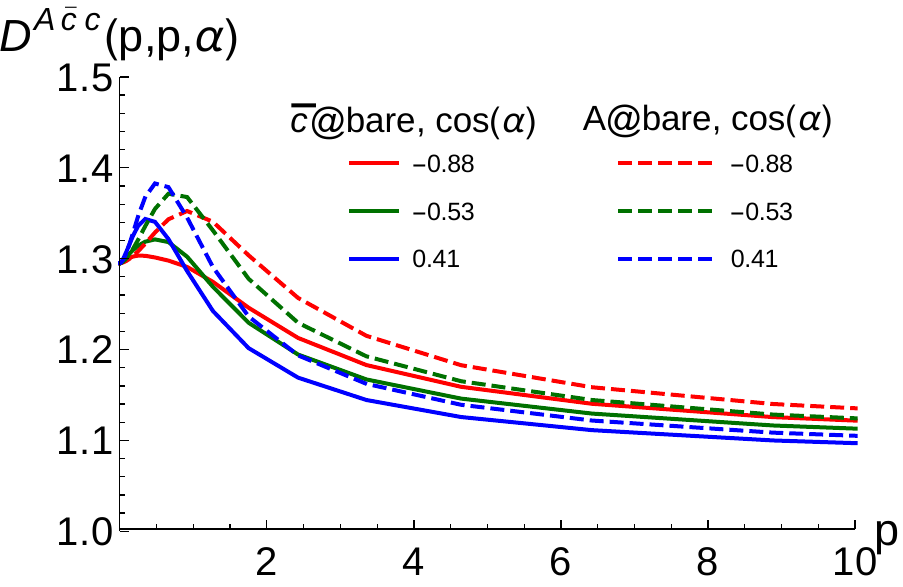}\\(d)}
\caption{\label{fig-ggv}(a) and (b): possible forms of the CRDSE for the ghost-gluon vertex.
(c) Result for the loop integrals in the ghost-gluon vertex DSE: the dashed lines represent the
lowest-order approximation, where all vertices are taken to be bare, while the continuous
lines are the result of the full calculation. (d) Dressing function of the ghost-gluon
vertex for equal ghost and antighost momentum and various angles as calculated from the
two different DSEs.}
\end{figure}

The truncated ghost-gluon vertex CRDSE is shown diagrammatically in Fig.~\ref{fig-ggv}a.
The first loop integral (the ``abelian diagram'') contains only ghost-gluon vertices, while
the second loop integral (the ``non-abelian diagram'') involves also a three-gluon vertex
(see later). The results \cite{Huber:2014isa} for the two diagrams are separately shown in Fig.~\ref{fig-ggv}c.
Figure~\ref{fig-ggv}b shows an alternative CRDSE, which differs from the one given in
Fig.~\ref{fig-ggv}a by the leg attached to the bare vertex (remember that in general a
$n$-point functions has $n$ possible DSEs). Because of truncation artefacts the two
equations do not yield exactly the same result. The differences depend on the specific
momentum configuration: in general they are small and they even practically vanish for
some momentum configurations, see Fig.~\ref{fig-ggv}d.

\begin{figure}[ht]
\centering
\includegraphics[width=.8\linewidth]{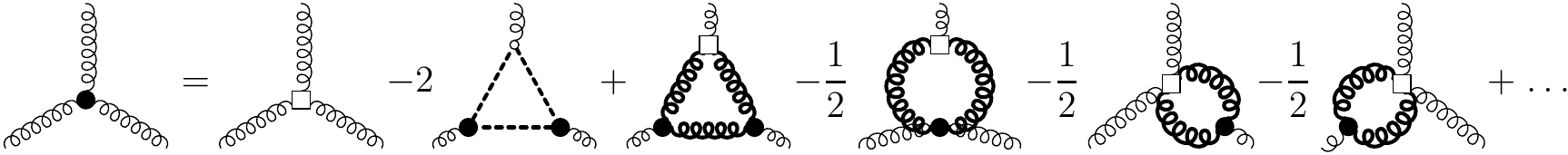}
\par\bigskip
\parbox{.35\linewidth}{\includegraphics[width=\linewidth]{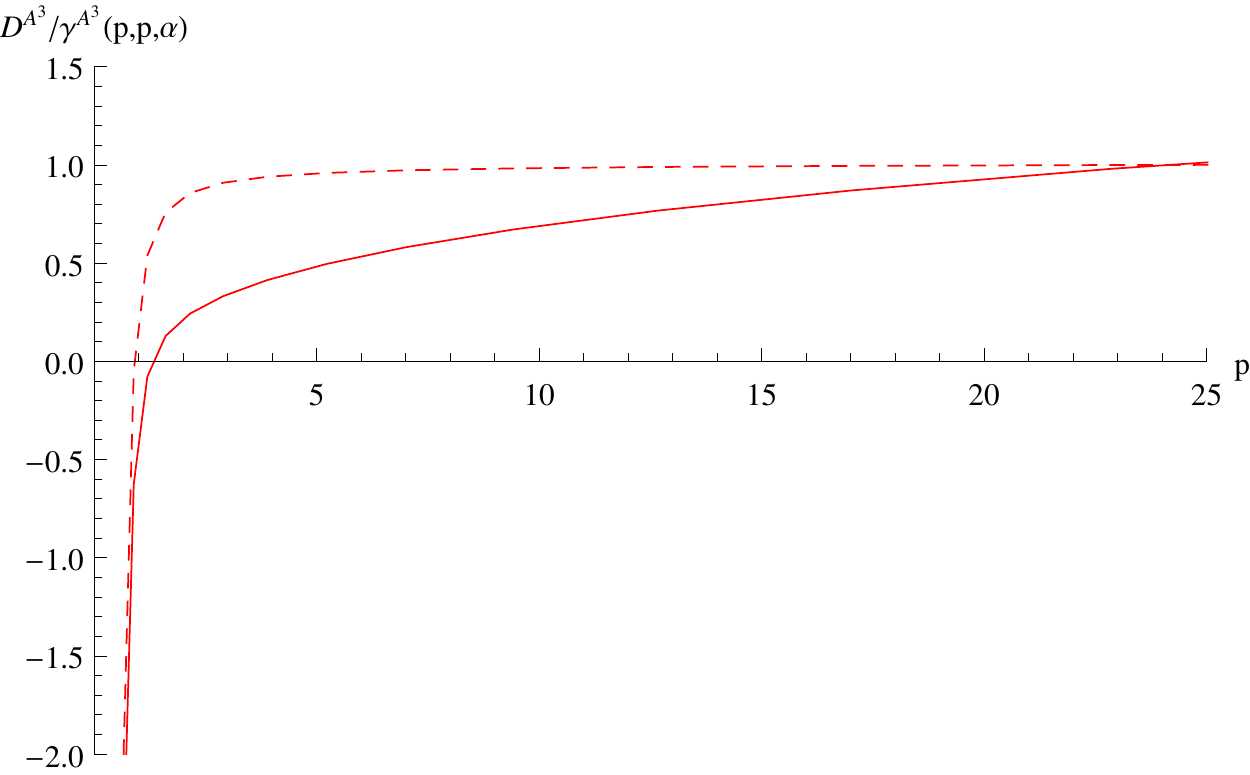}}\qquad\qquad
\parbox{.35\linewidth}{\includegraphics[width=\linewidth]{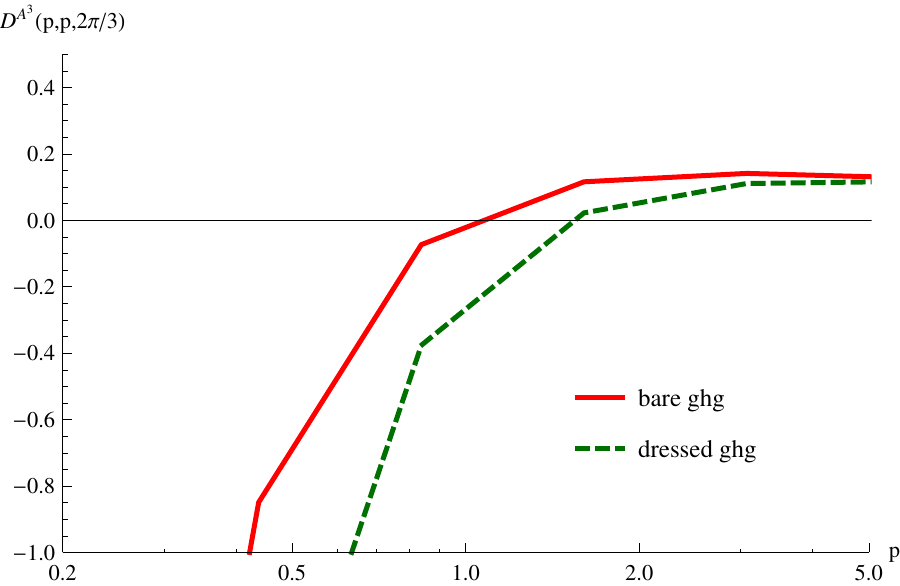}}
\caption{\label{fig-3gv}(top) Truncated CRDSE for the three-gluon vertex. (bottom left)
Dressing function of the perturbative tensor structure of the three gluon vertex at the
symmetric point: the dashed line is the result from the ghost triangle only, while the
continuous line is the result of the full calculation. (bottom right) Zoom around the
zero crossing showing the difference between a bare (continuous line) and dressed (dashed line)
ghost-gluon vertex.}
\end{figure}

Figure~\ref{fig-3gv} shows the truncated CRDSE for the three-gluon vertex. The full
four-gluon vertex is replaced by the bare one in the actual numerical calculation.
It turns out that the gluon triangle diagram is sub-leading in comparison to the ghost
triangle diagram; neglecting the swordfish diagrams altogether, on the other hand,
turns out to be too crude an approximation \cite{Huber:2014isa}.

\section{Quark Mass Function and Chiral Condensate}

The first calculations \cite{Pak:2011wu,Pak:2013uba} in the quark sector used an ansatz similar to
\Eqref{cx6} where, however, only the Dirac structure $\alpha_i$ was considered. The
resulting equations were plagued by linear divergences whose presence suggested that
this simple ansatz needed to be extended. While a full variational equation for the
quark-gluon kernel exists \cite{Campagnari:2015zsa} it is easier to motivate the presence
of the second Dirac structure in the following way: Consider first-order perturbation theory,
where the relevant part of the Hamilton operator has the form
\[
H = H_\mathrm{D} + g \hat\psi^\dag \valpha\cdot\vA \hat\psi ,
\]
with $H_\mathrm{D}$ being the Hamiltonian of the free Dirac theory. The first-order correction
$\ket{0}^{(1)}$ to the vacuum state $\ket{0}^{(0)}$ is given as usual by
\be\label{neu1}
\ket{0}^{(1)} = - \sum \frac{^{(0)}\bra{n} g \hat\psi^\dag \valpha\cdot\vA \hat\psi \ket{0}^{(0)}}{E_n} \, \ket{n}^{(0)}
\ee
where $\ket{n}^{(0)}$ are the excited eigenstates of the free theory. Working out the
matrix element \cite{Campagnari:2014hda} one finds (apart from numerical factors)
\[
\ket{0}^{(1)} \propto \hat\psi^\dag \valpha\cdot\vA \hat\psi \ket{0}^{(0)} .
\]
Once the coherent-state representation is used one arrives at a functional of the form of
\Eqref{cx6} with $s=0$ and $w=0$. The first term in \Eqref{cx6} is a BCS-type of wave functional,
already used in the Coulomb-gauge pairing model \cite{Finger:1981gm,Adler:1984ri,Alkofer:1988tc}.
If we use this ``vacuum'' (which, admittedly, is no eigenstate of $H_\mathrm{D}$) as bare
state and evaluate \Eqref{neu1}, we find a first-order correction coming with a Dirac
structure $\beta \alpha_i$. The inclusion of such a term in the ansatz is straightforward
\cite{Vastag:2015qjd,Campagnari:2016wlt}, and the resulting gap equation for the quark mass function
shows no spurious linear divergences any more. The resulting equation for the quark
mass function has the structure
\[
M(\vp) \sim \parbox[b]{4em}{\includegraphics[width=\linewidth]{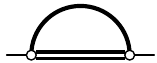}} + \parbox[b]{4em}{\includegraphics[width=\linewidth]{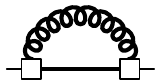}}
\]
where the first diagram involves a Coulomb-like interaction and the second one a gluon exchange.
Using for the Coulomb interaction [see second term in \Eqref{neu2} and \Eqref{neu3}]
the sum of a confining ($\propto1/p^4$, i.e.~linearly rising in coordinate space) and a
perturbative ($\sim1/p^2$) term
\be\label{neu4}
F(\vp) = \frac{8\pi \sigma_\mathrm{C}}{p^4} + \frac{\alpha}{p^2}
\ee
and taking the Coulomb string tension $\sigma_\mathrm{C}$ to be 2.5 times the Wilson string
tension we recover a chiral condensate of $\sim(-235\,\mathrm{MeV})^3$. Spontaneous breaking
of chiral symmetry breaking does not occur if the confining part of the non-abelian
Coulomb potential \eqref{neu4} is discarded.

\section{Conclusions}

We have presented some results of the CRDSE approach to treat non-Gaussian wave
functionals in the Hamiltonian formulation of a quantum field theory, presented first
in Ref.~\cite{Campagnari:2010wc} in the framework of pure Yang--Mills theory and
generalized to full QCD in Ref.~\cite{Campagnari:2015zsa}. In particular, we have shown
that a non-trivial Dirac structure is needed in order to avoid spurious linear divergences.
With the inclusion of the coupling of the quarks to the gluons in our vacuum wave functional
our approach is capable to reproduce the phenomenological value of the quark condensate.

\end{document}